\documentclass[twocolumn,aps,pra]{revtex4-2} %onecolumn
\usepackage{bm}    % bold math
\usepackage{amssymb,amsmath}
\usepackage{cases}
\usepackage{times}
\usepackage{graphicx}
\usepackage[colorlinks=true,linkcolor=blue,citecolor=blue,urlcolor=blue]{hyperref}

\begin{document}

\title{Size-dependent dynamical instability of periodic $\mathcal{PT}$-symmetric scattering systems}

\author{Chao Zheng}
\email{zhengchaonju@gmail.com}
\affiliation{School of Education, Jiangsu Open University, Nanjing 210036, China}

\begin{abstract}

While periodic $\mathcal{PT}$-symmetric structures offer a versatile platform for wave tailoring, their scattering responses are typically analyzed using stationary methods that presume dynamical stability. This assumption fails when time-growing bound states emerge, signaling a dynamical instability. Here, we analytically derive the instability threshold for a $\mathcal{PT}$-symmetric chain of $N$ unit cells with gain/loss strength $\gamma$. Our $S$-matrix analysis yields a closed-form threshold, $\gamma_c = 2\sin[\pi/(4N)]$, which scales as $\mathcal{O}(1/N)$ and vanishes in the thermodynamic limit. Consequently, enlarging such structures to access richer stationary band phenomena paradoxically triggers instability at weaker gain/loss. As confirmed by time-domain simulations, exceeding $\gamma_c$ causes exponentially growing bound states to overwhelm the system, rendering standard Bloch-wave descriptions physically irrelevant. Evaluated against this size-dependent threshold, many hallmarks of large $\mathcal{PT}$-symmetric structures, including gain-loss-induced localization, reflectionless transport, and coherent perfect absorbers and lasers, are found to lie within the dynamically unstable regime. Our findings thus establish that physical transport in non-Hermitian periodic systems is governed by a fundamental interplay between stationary band theory and finite-size stability limits.
\end{abstract}

\date{\today}

\maketitle

%%%%%%%%%%%%%%%%%%%%%%%%%%%%%%%%%%%%%%%%%%%%%%%%%%%%%%%%%%%%%%%%%%%%%%%%%%%%%%%%%%%%%%%%%%%%%%%%%%%%
\section{Introduction}
\label{sec:Introduction}
%%%%%%%%%%%%%%%%%%%%%%%%%%%%%%%%%%%%%%%%%%%%%%%%%%%%%%%%%%%%%%%%%%%%%%%%%%%%%%%%%%%%%%%%%%%%%%%%%%%%

Periodic media lie at the heart of wave physics.
Coherent scattering from a repeating unit organizes wave propagation into Bloch bands separated by forbidden gaps, a universal framework that governs electrons in crystals, light in photonic crystals, and sound in phononic structures~\cite{ashcroft1976,joannopoulos1997,hussein2014}.
While the wave dynamics of conservative, Hermitian systems rest on a mature, well-established paradigm, non-Hermiticity breaks free of these traditional constraints.
Non-Hermitian Hamiltonians describe open systems that exchange energy or particles with their environment, generically yielding complex spectra and nonorthogonal eigenstates~\cite{rotter2009,moiseyev2011}.
Far from merely complicating the mathematics, non-Hermiticity endows wave physics with fundamentally new degrees of freedom, offering a versatile platform for tailoring waves and enabling transport phenomena strictly prohibited in conservative settings~\cite{ashida2020,wang2023}.

A particularly important class of such open systems possesses parity-time ($\mathcal{PT}$) symmetry~\cite{konotop2016,el-ganainy2018,ozdemir2019,bender2024}, physically realized by a balanced spatial distribution of gain and loss.
Despite their open nature, $\mathcal{PT}$-symmetric Hamiltonians can exhibit entirely real spectra below a symmetry-breaking threshold~\cite{bender1998a}.
In scattering settings, this delicate balance gives rise to a variety of distinctive phenomena~\cite{makris2008,longhi2010,chong2011,lin2011,feng2013,fleury2014,garmon2015,zhu2016,shobe2021,lee2021a}, including double refraction~\cite{makris2008}, negative refraction~\cite{fleury2014}, unidirectional invisibility~\cite{lin2011,feng2013}, and coherent perfect absorbers and lasers (CPA lasers)~\cite{longhi2010,chong2011}.
Assembling $\mathcal{PT}$-symmetric cells into a periodic array further enriches this landscape.
As in ordinary crystals, wave propagation is governed by a Bloch phase and organized into bands~\cite{vazquez-candanedo2014,achilleos2017a,lee2023}.
Crucially, the number of unit cells $N$ provides a structural parameter for tailoring the scattering response through geometry alone.
Enlarging the structure is the standard recipe for sharpening the Bloch bands and generating additional Fabry-P\'erot transmission resonances and CPA-laser points~\cite{vazquez-candanedo2015,shramkova2016,achilleos2017a,ge2017,wu2019b,lee2023,perez-garrido2023}.
Periodic $\mathcal{PT}$-symmetric structures have thus become a canonical platform for non-Hermitian scattering, extensively investigated in photonics~\cite{vazquez-candanedo2014,vazquez-candanedo2015,shramkova2016,nguyen2016,achilleos2017a,ge2017,wu2019a,wu2019b,zheng2019,moreno-rodriguez2020,lee2023,perez-garrido2023,guo2023a,moreno-rodriguez2024,lee2024}, acoustics~\cite{zhu2014}, and electronic circuits~\cite{humire2019,lazo2023}.

These scattering properties are almost always obtained with time-independent methods, most commonly the transfer-matrix method~\cite{markos2008}, which extracts the reflection and transmission coefficients from the stationary scattering states.
Implicit in this approach is the assumption that the system is dynamically stable, so that a stationary description is meaningful at all.
This assumption is not automatic in non-Hermitian systems~\cite{zheng2025a}.
Just as an isolated $\mathcal{PT}$-symmetric Hamiltonian can cross a symmetry-breaking threshold beyond which complex eigenvalues render it dynamically unstable, an open $\mathcal{PT}$-symmetric chain coupled to leads can develop an instability of its own.
While the $\mathcal{PT}$-symmetry breaking of closed systems has been studied extensively~\cite{jin2009,joglekar2010,joglekar2011,barashenkov2013,agarwal2018}, the dynamical stability of open scattering systems has received far less attention~\cite{bendix2009,dmitriev2011}.
The distinction is essential: a closed system becomes unstable when a discrete eigenvalue acquires a positive imaginary part, whereas an open system supports a continuum of real-energy scattering states alongside its discrete states, and its stability is governed by the latter~\cite{zheng2025a}.
Determining whether, and at what gain/loss strength, such a system becomes unstable, and how this threshold scales with the system size $N$, is fundamental to understanding scattering in periodic non-Hermitian systems and sets the limits within which $\mathcal{PT}$-symmetric devices can operate as designed.

In this work, we analytically derive the instability threshold for a finite $\mathcal{PT}$-symmetric tight-binding chain composed of $N$ unit cells with balanced gain/loss strength $\gamma$. The instability is triggered by the emergence of time-growing bound states (TGBSs), which manifest as $S$-matrix poles in the first quadrant of the complex wave-number $k$ plane~\cite{zheng2025a}. By analyzing this pole structure, we show that the poles can reach the real axis only at the band center. This observation yields a closed-form expression for the instability threshold, $\gamma_c=2\sin[\pi/(4N)]$, the critical gain/loss strength at which the first TGBS emerges. For large $N$, the threshold scales as $\mathcal{O}(1/N)$ and vanishes in the thermodynamic limit. This inverse scaling reveals a fundamental tension intrinsic to periodic $\mathcal{PT}$-symmetric systems: although enlarging the structure is the standard route to sharpen the Bloch bands and enrich the stationary phenomenology, it paradoxically lowers $\gamma_c$ and drives the system unstable at progressively smaller gain/loss. Time-domain wave-packet simulations quantitatively confirm this boundary: below $\gamma_c$, the dynamics faithfully follow the stationary scattering predictions; at $\gamma_c$, the system emits persistent radiation; and above $\gamma_c$, the exponentially growing TGBS overwhelms the steady-state response. Evaluated against this size-dependent threshold, many hallmarks of large $\mathcal{PT}$-symmetric structures, including gain-loss-induced localization, reflectionless transport, and CPA lasers, are found to fall within the dynamically unstable regime. Our results establish that physical transport in periodic non-Hermitian systems is governed not by stationary band theory alone, but by a fundamental interplay between the Bloch phase and finite-size stability limits.

The paper is organized as follows.
Section~\ref{sec:model_and_method} introduces the periodic $\mathcal{PT}$-symmetric tight-binding model and the transfer-matrix method.
Section~\ref{sec:S_matrix_poles} analyzes the $S$-matrix pole structure, illustrates the pole migration for $N=3$, and analytically derives the instability threshold $\gamma_c(N)$ for arbitrary $N$.
Section~\ref{sec:time_dependent} verifies the threshold through time-dependent wave-packet simulations.
Section~\ref{sec:implications} re-interprets gain-loss-induced localization, reflectionless transport, and CPA lasers in light of the stability boundary.
Section~\ref{sec:conclusions} summarizes our findings.

%%%%%%%%%%%%%%%%%%%%%%%%%%%%%%%%%%%%%%%%%%%%%%%%%%%%%%%%%%%%%%%%%%%%%%%%%%%%%%%%%%%%%%%%%%%%%%%%%%%%
\section{Model and method}
\label{sec:model_and_method}
%%%%%%%%%%%%%%%%%%%%%%%%%%%%%%%%%%%%%%%%%%%%%%%%%%%%%%%%%%%%%%%%%%%%%%%%%%%%%%%%%%%%%%%%%%%%%%%%%%%%

%%%%%%%%%%%%%%%%%%%%%%%%%%%%%%%%%%%%%%%%%%%%%%%%%%%%%%%%%%%%%%%%%%%%%%%%%%%%%%%%%%%%%%%%%%%%%%%%%%%%
\subsection{Model}
\label{sec:model}
%%%%%%%%%%%%%%%%%%%%%%%%%%%%%%%%%%%%%%%%%%%%%%%%%%%%%%%%%%%%%%%%%%%%%%%%%%%%%%%%%%%%%%%%%%%%%%%%%%%%

\begin{figure}
  \centering
  \includegraphics[width=\columnwidth]{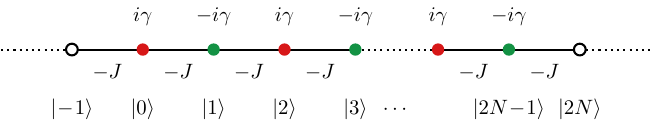}
  \caption{
   Schematic of the periodic $\mathcal{PT}$-symmetric scattering system. A scattering region of $N$ unit cells (sites $0$ to $2N{-}1$) is coupled to two semi-infinite Hermitian tight-binding leads. Each unit cell consists of a gain site ($+i\gamma$, red) and a loss site ($-i\gamma$, green).
    }
  \label{fig:Schematic_01}
\end{figure}

We consider a one-dimensional tight-binding chain consisting of a finite $\mathcal{PT}$-symmetric scattering region coupled to two semi-infinite Hermitian leads, as illustrated in Fig.~\ref{fig:Schematic_01}~\cite{vazquez-candanedo2014,vazquez-candanedo2015,moreno-rodriguez2020,lazo2023,moreno-rodriguez2024}.
This tight-binding model maps directly onto several experimental platforms, including arrays of coupled optical waveguides~\cite{guo2009,ruter2010,weimann2017}, photonic lattices~\cite{regensburger2012,regensburger2013}, and electrical circuits~\cite{schindler2011,liu2020b}, where balanced gain and loss are routinely engineered.
The Hamiltonian reads
\begin{equation}
  H = -J\sum_{j=-\infty}^{\infty}
  \left(|j\rangle\langle j+1| + \text{H.c.}\right)
  + \sum_{j=0}^{2N-1}\varepsilon_j|j\rangle\langle j|,
  \label{eq:Hamiltonian}
\end{equation}
where $|j\rangle$ is the Wannier state localized at site $j$ and $J$ is the nearest-neighbor hopping amplitude.
The on-site potentials are nonzero only within the scattering region, which spans sites $j=0,1,\ldots,2N{-}1$ and contains $N$ unit cells.
Each unit cell consists of a gain site followed by a loss site,
\begin{equation}
  \varepsilon_j = (-1)^j\, i\gamma, \quad j = 0, 1, \ldots, 2N-1,
  \label{eq:onsite_potential}
\end{equation}
with $\gamma > 0$ the balanced gain/loss strength.
The profile satisfies $\varepsilon_j = \varepsilon_{2N-1-j}^*$, ensuring $\mathcal{PT}$ symmetry of the scattering region, where $\mathcal{P}$ reflects the chain about its center and $\mathcal{T}$ denotes complex conjugation.
Throughout this work, we set $\hbar = J = a = 1$, with $a$ the lattice constant, so that energies are expressed in units of $J$, times in units of $\hbar/J$, and wave numbers in units of $1/a$.

%%%%%%%%%%%%%%%%%%%%%%%%%%%%%%%%%%%%%%%%%%%%%%%%%%%%%%%%%%%%%%%%%%%%%%%%%%%%%%%%%%%%%%%%%%%%%%%%%%%%
\subsection{Transfer-matrix method}
\label{sec:transfer_matrix}
%%%%%%%%%%%%%%%%%%%%%%%%%%%%%%%%%%%%%%%%%%%%%%%%%%%%%%%%%%%%%%%%%%%%%%%%%%%%%%%%%%%%%%%%%%%%%%%%%%%%

We analyze the system using the transfer-matrix method~\cite{markos2008}, a time-independent approach widely applied to $\mathcal{PT}$-symmetric scattering~\cite{chong2011,lin2011,feng2013,fleury2014,garmon2015,zhu2016,shobe2021,lee2021a,vazquez-candanedo2014,vazquez-candanedo2015,nguyen2016,shramkova2016,achilleos2017a,ge2017,wu2019a,wu2019b,zheng2019,moreno-rodriguez2020,lee2023,perez-garrido2023,guo2023a,lazo2023,lee2024,moreno-rodriguez2024}.
Conventionally, it constructs the stationary scattering states, the real-energy solutions of $H|\psi\rangle = E|\psi\rangle$ that satisfy scattering boundary conditions, and from them extracts the reflection and transmission coefficients.
Yet the same formalism also encodes the discrete spectrum of the open system: bound and resonant states manifest as poles of the $S$-matrix in the complex $k$ plane.
As we show in Sec.~\ref{sec:S_matrix_poles}, the distribution of these poles governs the dynamical stability of the system.
We therefore develop the stationary scattering description here, laying the groundwork for the stability analysis that follows.

Expanding the wave function in the Wannier basis, $|\psi\rangle = \sum_j \psi_j |j\rangle$, yields a recurrence relation for the amplitudes $\psi_j$,
\begin{equation}
  -\psi_{j-1} - \psi_{j+1} + \varepsilon_j \psi_j = E \psi_j.
  \label{eq:recurrence}
\end{equation}
In the leads ($j \leq -1$ and $j \geq 2N$), the on-site potential vanishes.
Equation~\eqref{eq:recurrence} then reduces to the relation for a free tight-binding chain, whose plane-wave solutions $e^{\pm ikj}$ obey the dispersion relation
\begin{equation}
  E = -2\cos k,
  \label{eq:dispersion}
\end{equation}
which defines the scattering continuum $E \in [-2,2]$ for $k \in (0,\pi)$.
A generalized scattering state thus asymptotically decomposes into incoming and outgoing plane waves,
\begin{equation}
  \psi_j =
  \begin{cases}
    A e^{ikj} + B e^{-ikj}, & j \leq -1, \\
    C e^{ikj} + D e^{-ikj}, & j \geq 2N,
  \end{cases}
  \label{eq:scattering_state}
\end{equation}
where $A$ and $D$ denote incoming amplitudes from the left and right leads, and $B$ and $C$ the corresponding outgoing amplitudes.

Inside the scattering region ($0 \leq j \leq 2N-1$), the recurrence relation~\eqref{eq:recurrence} can be recast in transfer-matrix form,
\begin{equation}
  \begin{pmatrix} \psi_{j+1} \\ \psi_j \end{pmatrix}
  = \mathcal{T}_j
  \begin{pmatrix} \psi_j \\ \psi_{j-1} \end{pmatrix},
  \quad
  \mathcal{T}_j =
  \begin{pmatrix} \varepsilon_j - E & -1 \\ 1 & 0 \end{pmatrix}.
  \label{eq:single_site_TM}
\end{equation}
Propagating across one unit cell (a gain site followed by a loss site) yields the unit-cell transfer matrix
\begin{equation}
  \mathcal{M} = \mathcal{T}_{\mathrm{loss}}\mathcal{T}_{\mathrm{gain}}
  =
  \begin{pmatrix}
    E^2+\gamma^2-1 & E+i\gamma \\
    -E+i\gamma     & -1
  \end{pmatrix},
  \label{eq:unit_cell_TM}
\end{equation}
which satisfies $\det\mathcal{M}=1$. Its eigenvalues can be parameterized as $\lambda_\pm = e^{\pm 2i\mu}$, where the Bloch-like index $\mu$ is fixed by $\cos(2\mu)=\tfrac{1}{2}\operatorname{Tr}\mathcal{M}$, giving the dispersion relation inside the scattering region~\cite{vazquez-candanedo2014},
\begin{equation}
  4\cos^2\mu = E^2 + \gamma^2.
  \label{eq:mu_dispersion}
\end{equation}
Depending on the incident energy and the gain/loss strength, $\mu$ transitions from purely real ($E^2+\gamma^2<4$) to purely imaginary ($E^2+\gamma^2>4$), separated by exceptional points at the band edges ($E^2+\gamma^2=4$).

Since all $N$ unit cells are identical and $\det\mathcal{M}=1$, the Chebyshev identity gives the $N$-cell transfer matrix in closed form~\cite{markos2008,achilleos2017a}:
\begin{equation}
  \mathcal{M}^N
  = \mathcal{M}\frac{\sin(2N\mu)}{\sin(2\mu)}
  - \mathbf{I}\frac{\sin[2(N-1)\mu]}{\sin(2\mu)}.
  \label{eq:chebyshev}
\end{equation}
The matrix $\mathcal{M}^N$ maps consecutive site amplitudes $(\psi_0,\psi_{-1})^T$ to $(\psi_{2N},\psi_{2N-1})^T$.
To express the scattering problem in terms of the plane-wave amplitudes in Eq.~\eqref{eq:scattering_state}, we perform the basis transformation~\cite{markos2008}
\begin{equation}
  M = Q^{-1}\mathcal{M}^N Q,
  \quad
  Q =
  \begin{pmatrix}1 & 1\\ e^{-ik} & e^{ik}\end{pmatrix},
  \label{eq:M_from_Mcal}
\end{equation}
so that
\begin{equation}
  \begin{pmatrix}
    Ce^{2iNk} \\ De^{-2iNk}
  \end{pmatrix}
  = M
  \begin{pmatrix}
    A \\ B
  \end{pmatrix}.
  \label{eq:transfer_matrix_def}
\end{equation}
Imposing the boundary condition for left incidence ($D = 0$) or right incidence ($A = 0$) in Eq.~\eqref{eq:transfer_matrix_def} yields the scattering amplitudes in terms of the transfer-matrix elements:
\begin{equation}
  t = \frac{1}{M_{22}}, \quad
  r_L = -\frac{M_{21}}{M_{22}}, \quad
  r_R = \frac{M_{12}}{M_{22}}.
  \label{eq:scattering_from_M}
\end{equation}
Here $r_L$ ($r_R$) is the reflection amplitude for incidence from the left (right), and the transmission amplitude $t$ is the same for both incidence directions by virtue of $\det M = 1$.
The corresponding transmission and reflection coefficients are $T=|t|^{2}$ and $R_{L(R)}=|r_{L(R)}|^{2}$, respectively.

Carrying out the matrix multiplication in Eq.~\eqref{eq:M_from_Mcal} and simplifying using the dispersion relation~\eqref{eq:mu_dispersion}, one obtains
\begin{align}
  M_{11}
  &= \cos(2N\mu)
   + i\cot k\tan\mu\sin(2N\mu), \label{eq:M11}\\
  M_{12}
  &= \frac{i\gamma e^{ik}(2\sin k - \gamma)}{2\sin k}\frac{\sin(2N\mu)}{\sin(2\mu)}, \label{eq:M12}\\
  M_{21}
  &= \frac{i\gamma e^{-ik}(2\sin k + \gamma)}{2\sin k}\frac{\sin(2N\mu)}{\sin(2\mu)}, \label{eq:M21}\\
  M_{22}
  &= \cos(2N\mu)
   - i\cot k\tan\mu\sin(2N\mu). \label{eq:M22}
\end{align}
From these expressions one finds $M_{11} = M_{22}^*$~\cite{longhi2010}, which, together with $\det M = 1$, yields the generalized conservation relation for $\mathcal{PT}$-symmetric scattering~\cite{ge2012},
\begin{equation}
  |T - 1| = \sqrt{R_LR_R}.
  \label{eq:generalized_conservation}
\end{equation}
The transmission coefficient $T = |t|^2 = 1/|M_{22}|^2$ is thus obtained analytically as
\begin{equation}
  T
  = \frac{1}{
    1 - \dfrac{\gamma^2\sin^2(2N\mu)}
              {4\sin^2 k\cos^2\mu}
  },
  \label{eq:T_exact}
\end{equation}
in agreement with Refs.~\cite{vazquez-candanedo2014,vazquez-candanedo2015,moreno-rodriguez2020,moreno-rodriguez2024}.
Equations~\eqref{eq:scattering_from_M}--\eqref{eq:T_exact} constitute the stationary scattering description of the finite periodic $\mathcal{PT}$-symmetric chain.

%%%%%%%%%%%%%%%%%%%%%%%%%%%%%%%%%%%%%%%%%%%%%%%%%%%%%%%%%%%%%%%%%%%%%%%%%%%%%%%%%%%%%%%%%%%%%%%%%%%%
\section{$S$-matrix poles and the onset of dynamical instability}
\label{sec:S_matrix_poles}
%%%%%%%%%%%%%%%%%%%%%%%%%%%%%%%%%%%%%%%%%%%%%%%%%%%%%%%%%%%%%%%%%%%%%%%%%%%%%%%%%%%%%%%%%%%%%%%%%%%%

The stationary formalism of Sec.~\ref{sec:model_and_method} yields reflection and transmission coefficients for arbitrary system size $N$ and gain/loss strength $\gamma$.
Whether these coefficients describe the actual time evolution of a wave packet, however, depends on the dynamical stability of the system.
We now establish this stability criterion and show that it is governed by the distribution of $S$-matrix poles~\cite{zheng2025a}.

The spectrum of the open system has two components with distinct dynamical roles.
The continuum of scattering states carries real energies $E = -2\cos k$ and evolves through bounded oscillatory phases $e^{-iEt}$; it is inherently stable.
Instability can originate only in the discrete spectrum, from bound states that acquire complex energies.
An incident wave packet, even when prepared far from the scattering region, inevitably retains nonzero overlap with the evanescent tails of such states.
If any bound-state energy satisfies $\mathrm{Im}\,E > 0$, this overlap is exponentially amplified, giving rise to a TGBS that rapidly overwhelms the scattering dynamics and renders the stationary coefficients unphysical.
Dynamical stability is therefore determined entirely by the discrete spectrum.

These discrete states are encoded in the poles of the $S$ matrix~\cite{moiseyev2011,sasada2011,hatano2014,garmon2015}.
Since all scattering amplitudes in Eq.~\eqref{eq:scattering_from_M} share the common denominator $M_{22}$, the pole condition $M_{22}=0$ reads
\begin{equation}
  \cos(2N\mu) - i\cot k\tan\mu\sin(2N\mu) = 0.
  \label{eq:pole_condition}
\end{equation}
Its solutions form a discrete set of complex wave numbers $k_n = k_n^r + i k_n^i$ with energies $E_n = -2\cos k_n$.
The corresponding eigenstates are Siegert states~\cite{siegert1939}, which satisfy purely outgoing boundary conditions:
\begin{equation}
  \psi_{n,j}^{S}(t) = e^{-iE_n^r t} e^{E_n^i t}
  \times
  \begin{cases}
    Be^{-ik_n^r j}e^{k_n^i j}, & j \leq -1, \\
    Ce^{ik_n^r j}e^{-k_n^i j}, & j \geq 2N,
  \end{cases}
  \label{eq:Siegert_state}
\end{equation}
with $E_n^i = 2\sin k_n^r \sinh k_n^i$.

The position of each pole in the complex $k$ plane fixes the physical nature of its Siegert state~\cite{sasada2011,hatano2014,garmon2015}.
A first-quadrant pole ($k_n^r > 0$, $k_n^i > 0$) has $E_n^i > 0$: it describes a spatially localized, temporally growing state, i.e., a TGBS, whose presence renders the system dynamically unstable.
Second-quadrant poles correspond to time-decaying bound states, and lower-half-plane poles to scattering resonances; both leave the system stable.
A pole on the real axis is a spectral singularity~\cite{mostafazadeh2009,longhi2009,ramezani2014,wang2016,jin2018b}, at which the scattering amplitudes diverge and the Siegert state acquires a real energy.
On the positive real axis, this state carries purely outgoing waves and describes self-sustained lasing;
on the negative real axis, it carries purely incoming waves and describes coherent perfect absorption~\cite{chong2010,li2017a,baranov2017}.

In Hermitian systems, upper-half-plane poles are confined to the imaginary axis and the Brillouin-zone boundaries ($\mathrm{Re}\,k=0$ or $\pi$), and real-axis poles are forbidden~\cite{garmon2015}, so no TGBS can occur and the system is always stable.
Non-Hermiticity lifts these restrictions, allowing poles to migrate onto the real axis as spectral singularities or cross into the first quadrant as TGBSs.
The first pole to enter the first quadrant therefore marks the stability boundary of the open $\mathcal{PT}$-symmetric chain.
In what follows, we trace this pole migration numerically for $N=3$ (Sec.~\ref{sec:pole_N3}), then derive the instability threshold $\gamma_c(N)$ in closed form for arbitrary $N$ (Sec.~\ref{sec:gamma_c}).

%%%%%%%%%%%%%%%%%%%%%%%%%%%%%%%%%%%%%%%%%%%%%%%%%%%%%%%%%%%%%%%%%%%%%%%%%%%%%%%%%%%%%%%%%%%%%%%%%%%%
\subsection{Pole distribution for $N=3$}
\label{sec:pole_N3}
%%%%%%%%%%%%%%%%%%%%%%%%%%%%%%%%%%%%%%%%%%%%%%%%%%%%%%%%%%%%%%%%%%%%%%%%%%%%%%%%%%%%%%%%%%%%%%%%%%%%

\begin{figure*}
\centering
\includegraphics[width=2\columnwidth]{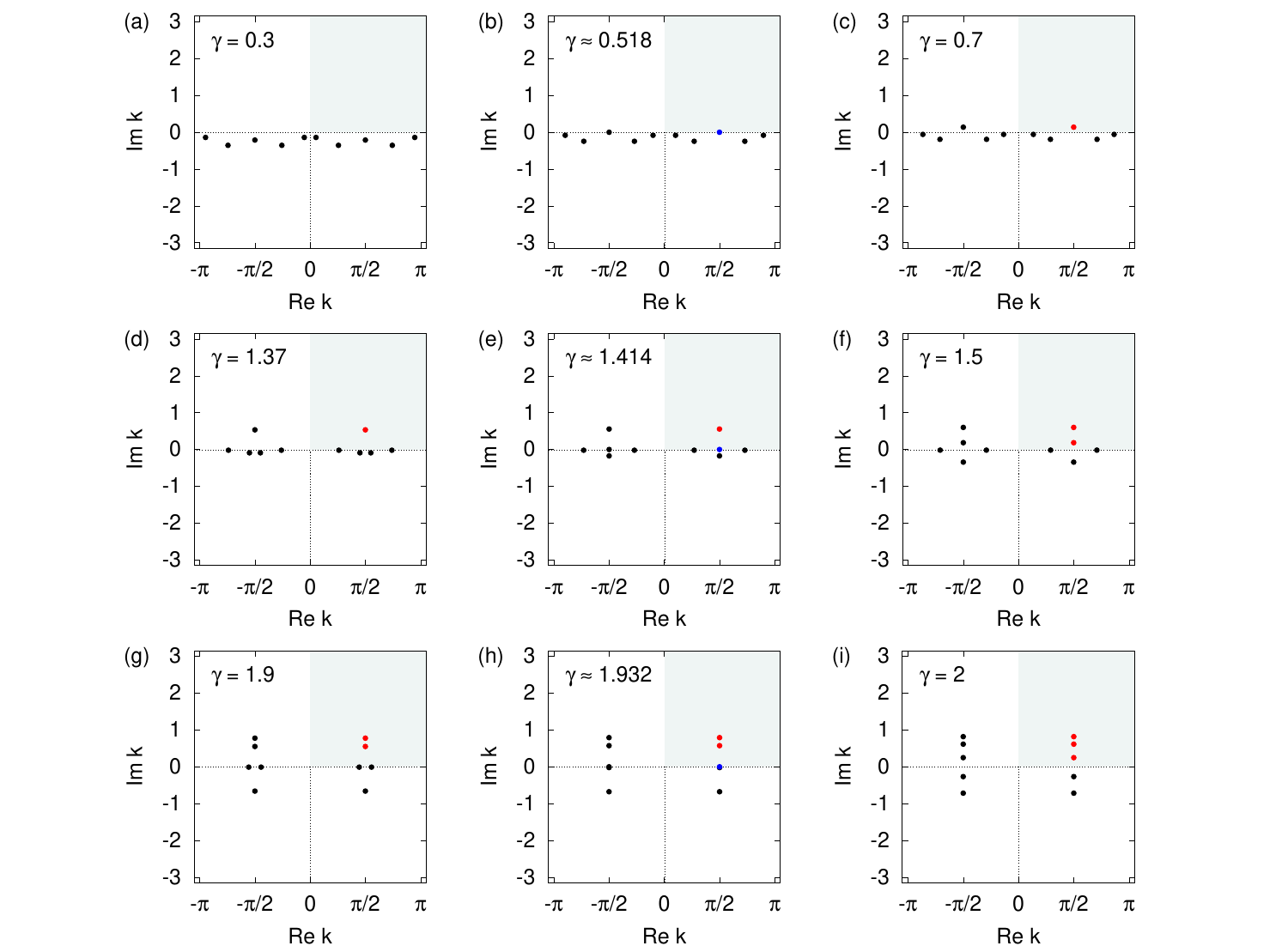}
\caption{
$S$-matrix poles in the complex $k$ plane for $N=3$ at nine values of $\gamma$:
(a) $\gamma=0.3$,
(b) $\gamma=\gamma_c=(\sqrt{6}-\sqrt{2})/2\approx0.518$,
(c) $\gamma=0.7$,
(d) $\gamma=1.37$,
(e) $\gamma = \sqrt{2}\approx1.414$,
(f) $\gamma=1.5$,
(g) $\gamma=1.9$,
(h) $\gamma=(\sqrt{6}+\sqrt{2})/2\approx1.932$, and
(i) $\gamma=2$.
Blue points on the positive real axis denote spectral singularities; red points in the first quadrant (shaded) denote TGBSs.
Beyond $\gamma_c$ [panel~(b)], poles enter the first quadrant, so the system hosts a TGBS and becomes dynamically unstable.
}
\label{fig:01_S_matrix_poles}
\end{figure*}

\begin{figure}
\centering
\includegraphics[width=\columnwidth]{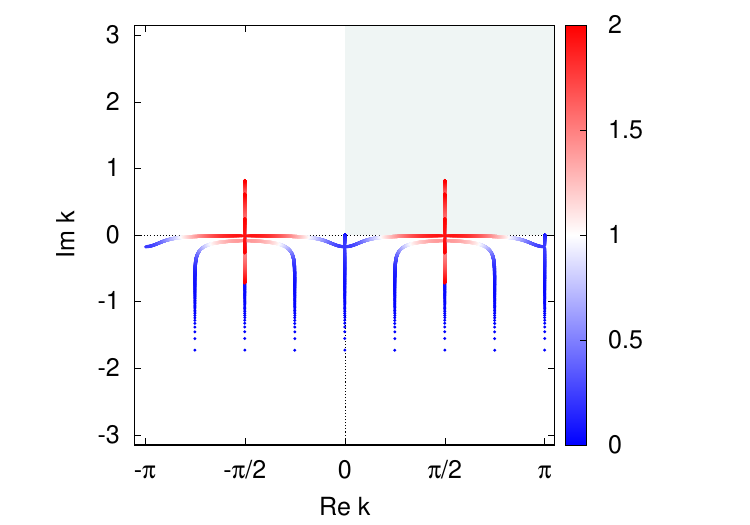}
\caption{
Trajectories of the $S$-matrix poles in the complex $k$ plane as a function of $\gamma$ for $N=3$.
The color scale indicates the value of $\gamma$, ranging from $0$ (blue) to $2$ (red).
Every pole crosses the real axis at $\mathrm{Re}\,k=\pm\pi/2$.
}
\label{fig:02_S_matrix_poles_trajectory}
\end{figure}

Figure~\ref{fig:01_S_matrix_poles} displays the $S$-matrix pole distribution for $N=3$ at nine representative values of $\gamma$, with spectral singularities (poles on the positive real axis) marked in blue and TGBSs (first-quadrant poles) marked in red.
Figure~\ref{fig:02_S_matrix_poles_trajectory} shows the full pole trajectories as $\gamma$ varies from $0$ to $2$.

At weak gain/loss [$\gamma=0.3$, Fig.~\ref{fig:01_S_matrix_poles}(a)], all poles lie in the lower half of the complex $k$ plane and correspond to resonant and antiresonant states, so the system is dynamically stable.
$\mathcal{PT}$ symmetry pairs poles at $(k_n, -k_n^*)$~\cite{simon2019}, so the distribution is symmetric about the imaginary axis.
Among the five poles with $\mathrm{Re}\,k>0$, one lies on the line $\mathrm{Re}\,k=\pi/2$, and the remaining four form an inner pair and an outer pair arranged symmetrically about this line.

As $\gamma$ increases, the pole on $\mathrm{Re}\,k = \pi/2$ moves upward and is the first to reach the real axis, becoming a spectral singularity at $\gamma_c=(\sqrt{6}-\sqrt{2})/2$ [Fig.~\ref{fig:01_S_matrix_poles}(b)].
This value defines the instability threshold for $N=3$.
A further small increase of $\gamma$ pushes the pole into the first quadrant [Fig.~\ref{fig:01_S_matrix_poles}(c)], producing a TGBS and rendering the system dynamically unstable.
The remaining two pairs follow the same pattern: for each pair, the two poles converge toward $\mathrm{Re}\,k=\pi/2$, coalesce, and then split along the imaginary direction, with the ascending member crossing the real axis as a spectral singularity before entering the first quadrant as an additional TGBS.
The inner pair crosses at $\gamma=\sqrt{2}$ [Fig.~\ref{fig:01_S_matrix_poles}(e)] and the outer pair at $\gamma=(\sqrt{6}+\sqrt{2})/2$ [Fig.~\ref{fig:01_S_matrix_poles}(h)].
By $\gamma=2$ [Fig.~\ref{fig:01_S_matrix_poles}(i)], all five poles with $\mathrm{Re}\,k>0$ have collapsed onto $\mathrm{Re}\,k=\pi/2$, three of them inside the first quadrant, so the system hosts three TGBSs.

A salient feature of Fig.~\ref{fig:02_S_matrix_poles_trajectory} is that every pole crosses the real axis at $\mathrm{Re}\,k=\pm\pi/2$.
We show in the next subsection that this is a general property of the periodic $\mathcal{PT}$-symmetric chain, and obtain the corresponding $\gamma$ values in closed form.

%%%%%%%%%%%%%%%%%%%%%%%%%%%%%%%%%%%%%%%%%%%%%%%%%%%%%%%%%%%%%%%%%%%%%%%%%%%%%%%%%%%%%%%%%%%%%%%%%%%%
\subsection{Instability threshold for general $N$}
\label{sec:gamma_c}
%%%%%%%%%%%%%%%%%%%%%%%%%%%%%%%%%%%%%%%%%%%%%%%%%%%%%%%%%%%%%%%%%%%%%%%%%%%%%%%%%%%%%%%%%%%%%%%%%%%%

The $N=3$ analysis showed that every pole with $\mathrm{Re}\,k>0$ enters the first quadrant by first crossing the real axis as a spectral singularity.
The instability threshold $\gamma_c$ is therefore the smallest $\gamma$ at which the pole condition~\eqref{eq:pole_condition} admits a real solution $k$.

For $k \in (0, \pi)$, the dispersion relation~\eqref{eq:mu_dispersion} reads $4\cos^{2}\mu = 4\cos^{2}k+\gamma^{2}$, so $\mu$ is either real (for $4\cos^{2}k + \gamma^{2} \leq 4$) or purely imaginary (for $4\cos^{2}k + \gamma^{2} > 4$).
The imaginary branch is excluded: substituting $\mu = i\phi$ with $\phi > 0$ into Eq.~\eqref{eq:M22} gives $\mathrm{Re}\,M_{22}=\cosh(2N\phi)>0$, so $M_{22}$ cannot vanish.
For real $\mu$, the real and imaginary parts of $M_{22}$ must vanish independently:
\begin{align}
  \cos(2N\mu) &= 0,
  \label{eq:pole_condition_Re} \\
  \cot k \tan\mu \sin(2N\mu) &= 0.
  \label{eq:pole_condition_Im}
\end{align}
Equation~\eqref{eq:pole_condition_Re} forces $\sin(2N\mu) \neq 0$ and $\tan\mu \neq 0$, so Eq.~\eqref{eq:pole_condition_Im} reduces to $\cot k = 0$, i.e.,
\begin{equation}
  k = \frac{\pi}{2}.
  \label{eq:k_pi_over_2}
\end{equation}
Spectral singularities of the periodic $\mathcal{PT}$-symmetric chain thus occur exclusively at $k=\pi/2$ (equivalently, at the band center $E=0$), consistent with the pole trajectories in Fig.~\ref{fig:02_S_matrix_poles_trajectory}.

Substituting $k = \pi/2$ into Eq.~\eqref{eq:mu_dispersion} gives $\gamma = 2\cos\mu$.
On the branch $\mu \in (0, \pi/2)$, Eq.~\eqref{eq:pole_condition_Re} admits $N$ solutions,
\begin{equation}
  \mu_n = \frac{(2n+1)\pi}{4N},
  \quad n = 0,1,\ldots,N{-}1,
  \label{eq:mu_n}
\end{equation}
with corresponding gain/loss strengths
\begin{equation}
  \gamma_n
  = 2\cos\left[\frac{(2n+1)\pi}{4N}\right],
  \quad n = 0,1,\ldots,N{-}1.
  \label{eq:gamma_n}
\end{equation}
The sequence $\gamma_0 > \gamma_1 > \cdots > \gamma_{N-1}$ enumerates the $N$ successive values at which poles cross the real axis.
For $N = 3$, Eq.~\eqref{eq:gamma_n} yields $\gamma_0 = 2\cos(\pi/12) = (\sqrt{6}+\sqrt{2})/2$, $\gamma_1 = 2\cos(\pi/4) = \sqrt{2}$, and $\gamma_2 = 2\cos(5\pi/12) = (\sqrt{6}-\sqrt{2})/2$, in agreement with the numerical values in Fig.~\ref{fig:01_S_matrix_poles}.

The onset of instability is set by the smallest member of this sequence,
\begin{equation}
  \gamma_c = \gamma_{N-1} = 2\sin\!\left(\frac{\pi}{4N}\right).
  \label{eq:gamma_c}
\end{equation}
For $\gamma<\gamma_c$, no pole lies in the first quadrant, and the system is dynamically stable; for $\gamma>\gamma_c$, at least one TGBS is present, and the system is unstable.

\begin{figure}
\centering
\includegraphics[width=\columnwidth]{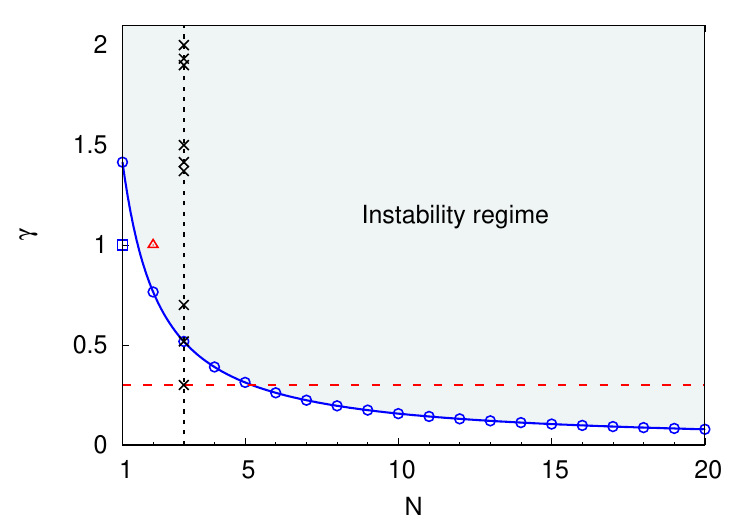}
\caption{
Instability threshold $\gamma_c = 2\sin[\pi/(4N)]$ (solid curve) as a function of system size $N$.
In the shaded region above the curve, at least one TGBS exists and the system is dynamically unstable.
The black crosses at $N=3$ mark the nine $\gamma$ values used in Fig.~\ref{fig:01_S_matrix_poles}.
The dashed lines, the blue square, and the red triangle indicate parameter values analyzed in Sec.~\ref{sec:implications}.
}
\label{fig:03_onset_condition}
\end{figure}

Equation~\eqref{eq:gamma_c} is the central result of this work: the dynamical stability of the periodic $\mathcal{PT}$-symmetric chain is controlled by its size.
The single-cell case $N=1$ gives $\gamma_c=\sqrt{2}$, recovering the spectral-singularity condition for the $\mathcal{PT}$-symmetric dimer coupled to leads~\cite{zhu2016}.
In the large-$N$ limit,
\begin{equation}
  \gamma_c \approx \frac{\pi}{2N},
  \label{eq:gamma_c_large_N}
\end{equation}
so the threshold decays inversely with system size and vanishes in the thermodynamic limit.
Figure~\ref{fig:03_onset_condition} shows $\gamma_c(N)$ in the $(\gamma, N)$ plane.
The black crosses at $N=3$ mark the nine $\gamma$ values of Fig.~\ref{fig:01_S_matrix_poles}; only the first two [panels~(a) and~(b)] lie within or on the boundary of the stable regime.

The $1/N$ scaling of the threshold carries a clear physical message.
The stationary scattering picture rests on the system being dynamically stable; the moment a TGBS appears, its exponential growth overwhelms the scattering-state contributions, and the dynamics is no longer captured by the stationary states.
Equation~\eqref{eq:gamma_c_large_N} shows that this happens at progressively weaker gain and loss as the structure grows.
Periodic $\mathcal{PT}$-symmetric chains are commonly studied for tens or hundreds of unit cells at gain/loss strengths of order unity~\cite{vazquez-candanedo2014,vazquez-candanedo2015,lazo2023,moreno-rodriguez2024}.
Even for a moderate chain length of $N=10$, the threshold is only $\gamma_c \approx 0.16$, so a system at $\gamma$ of order unity sits well inside the dynamically unstable regime.
For large structures, the stationary scattering picture describes the wave-packet dynamics only within a narrow window near $\gamma = 0$, and outside it the conventional scattering picture must be applied with care.

%%%%%%%%%%%%%%%%%%%%%%%%%%%%%%%%%%%%%%%%%%%%%%%%%%%%%%%%%%%%%%%%%%%%%%%%%%%%%%%%%%%%%%%%%%%%%%%%%%%%
\section{Time-dependent verification}
\label{sec:time_dependent}
%%%%%%%%%%%%%%%%%%%%%%%%%%%%%%%%%%%%%%%%%%%%%%%%%%%%%%%%%%%%%%%%%%%%%%%%%%%%%%%%%%%%%%%%%%%%%%%%%%%%

The preceding analysis predicts a sharp stability boundary at $\gamma_c = 2\sin[\pi/(4N)]$: below it the system is dynamically stable; above it at least one TGBS makes it unstable.
We now test this prediction directly by evolving wave packets in time for $N=3$ at the three representative gain/loss strengths of Fig.~\ref{fig:01_S_matrix_poles}: (i)~$\gamma=0.3 < \gamma_c$ (dynamically stable), (ii)~$\gamma=\gamma_c=(\sqrt{6}-\sqrt{2})/2$ (spectral singularity), and (iii)~$\gamma=0.7 > \gamma_c$ (dynamically unstable).

\begin{figure*}
\centering
\includegraphics[width=2\columnwidth]{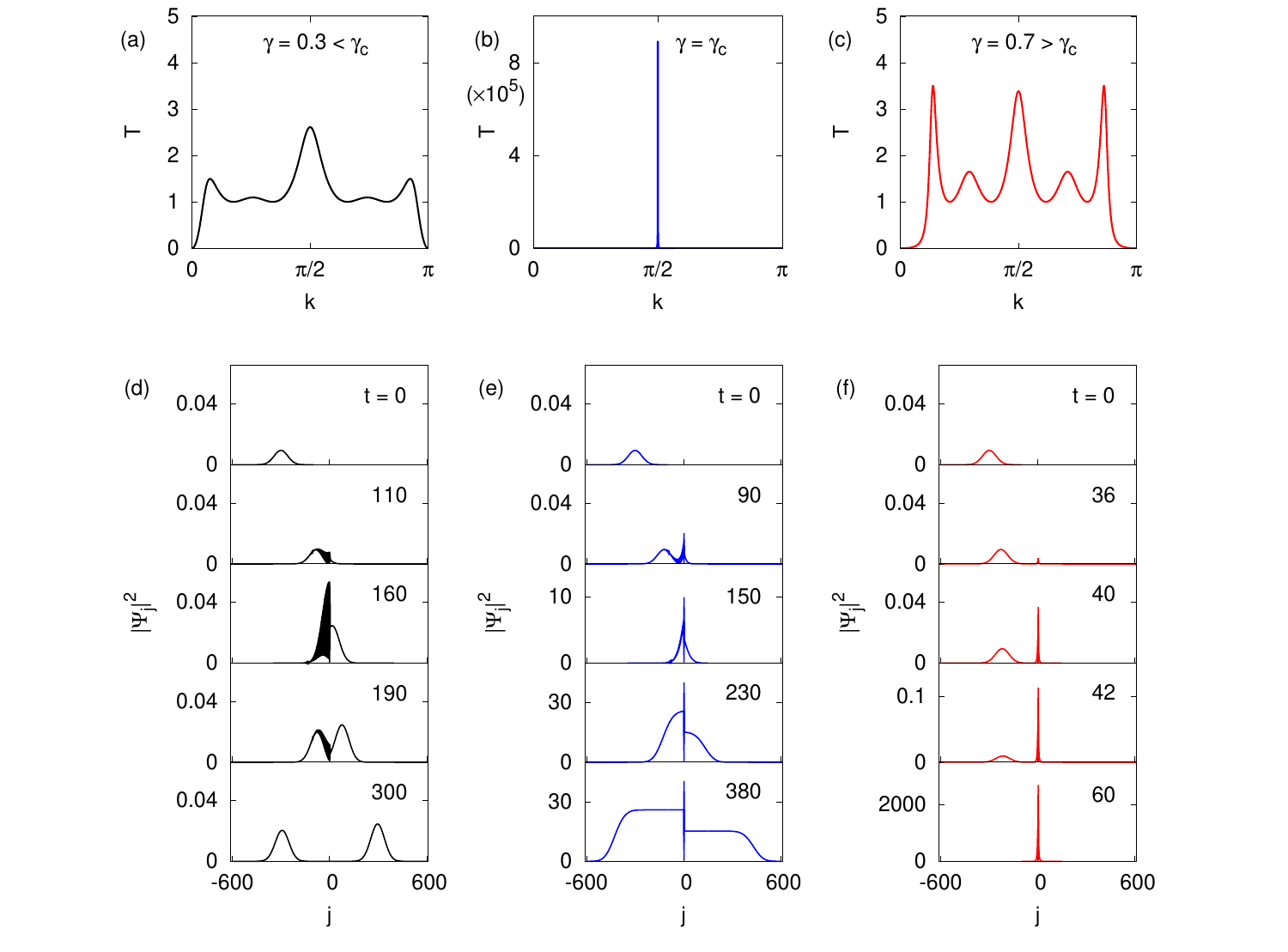}
\caption{
Time-dependent wave-packet simulations for $N = 3$ at three representative values of $\gamma$ from Fig.~\ref{fig:01_S_matrix_poles}:
(a,d)~$\gamma = 0.3 < \gamma_c$ (dynamically stable),
(b,e)~$\gamma = \gamma_c = (\sqrt{6}-\sqrt{2})/2$ (spectral singularity), and
(c,f)~$\gamma = 0.7 > \gamma_c$ (dynamically unstable).
Top row [(a)--(c)]: transmission coefficient $T(k)$ from the stationary calculation [Eq.~\eqref{eq:T_exact}].
Bottom row [(d)--(f)]: snapshots of the intensity profile $|\Psi_j(t)|^2$ at selected times.
The simulations use a Gaussian wave packet [Eq.~\eqref{eq:initial_wave_packet}] with $k_0 = \pi/2$, $\sigma = 60$, $j_0 = -300$, on a lattice of $L = 1200$ sites.
}
\label{fig:04_time_dependent_results}
\end{figure*}

The simulations use a finite lattice of $L = 1200$ sites, consisting of a central scattering region of $2N = 6$ sites connected to two truncated leads of $597$ sites each.
The initial state is a Gaussian wave packet
\begin{equation}
|\Psi(0)\rangle = \mathcal{N}^{-1} \sum_j e^{-(j-j_0)^2/2\sigma^2} e^{ik_0 j} |j\rangle,
\label{eq:initial_wave_packet}
\end{equation}
centered at site $j_0 = -300$ in the left lead, with half-width $\sigma = 60$ and central wave number $k_0 = \pi/2$; the normalization constant $\mathcal{N}$ ensures $\sum_j |\Psi_j(0)|^2 = 1$.
The time evolution $|\Psi(t)\rangle = e^{-iHt}|\Psi(0)\rangle$ is computed by exact diagonalization of the $L\times L$ Hamiltonian~\cite{zheng2025a}.
At $k_0 = \pi/2$, the group velocity is $v_g = 2\sin k_0 = 2$, so the wave packet reaches the scattering region at $t \approx |j_0|/v_g = 150$.
Figure~\ref{fig:04_time_dependent_results} compares the stationary transmission $T(k)$ [Eq.~\eqref{eq:T_exact}] in the top row with snapshots of the intensity profile $|\Psi_j(t)|^2$ in the bottom row.

Below the threshold ($\gamma = 0.3$), all $S$-matrix poles lie in the lower half of the complex $k$ plane [Fig.~\ref{fig:01_S_matrix_poles}(a)] and the system is dynamically stable.
The transmission $T(k)$ [Fig.~\ref{fig:04_time_dependent_results}(a)] is smooth and finite, with each pole at $\mathrm{Re}\,k > 0$ producing a Breit-Wigner-type resonance; the real and imaginary parts of the pole location set the resonance position and linewidth, respectively.
The wave-packet simulation [Fig.~\ref{fig:04_time_dependent_results}(d)] exhibits ordinary scattering: the incident packet propagates toward the scattering region, interacts with it, and splits into reflected and transmitted components.
The transmitted intensity at $t = 300$ is $\sum_{j \geq 2N} |\Psi_j|^2 = 2.60$, in close agreement with the stationary prediction $T(\pi/2) = 2.61$.
In the stable regime, the stationary scattering coefficients faithfully describe the dynamics.

At the threshold ($\gamma = \gamma_c$), a pole reaches the real axis at $k = \pi/2$ [Fig.~\ref{fig:01_S_matrix_poles}(b)], producing a spectral singularity at which $T(k)$ diverges [Fig.~\ref{fig:04_time_dependent_results}(b)].
The time-dependent simulation [Fig.~\ref{fig:04_time_dependent_results}(e)] reveals the corresponding lasing behavior: once the incident packet reaches the scattering region ($t \gtrsim 150$), persistent outgoing waves of constant amplitude radiate from the scattering center in both directions~\cite{wang2016,jin2018}.
The threshold thus marks the onset of self-sustained emission, the boundary between stable and unstable dynamics.

Above the threshold ($\gamma = 0.7$), a pole has crossed into the first quadrant at $k = 1.571 + 0.140i$ [Fig.~\ref{fig:01_S_matrix_poles}(c)], corresponding to a TGBS with complex energy $E = 0.280i$, and the system is dynamically unstable.
The stationary $T(k)$ remains smooth and finite [Fig.~\ref{fig:04_time_dependent_results}(c)] and carries no trace of this qualitative change in the pole structure.
The wave-packet dynamics, however, reveals drastically different behavior [Fig.~\ref{fig:04_time_dependent_results}(f)].
Well before the packet reaches the scattering region, a localized peak emerges at the system center whose intensity grows exponentially at the rate $2\,\mathrm{Im}\,E = 0.560$.
Although the initial packet is centered $300$ sites away from the scattering region, its nonzero overlap with the evanescent tails of the TGBS is exponentially amplified, and by $t = 60$ this growing component already overwhelms any reflected or transmitted signal.
The system is dynamically unstable, and the stationary scattering coefficients no longer describe its evolution.

These three cases confirm that $\gamma_c = 2\sin[\pi/(4N)]$, derived in Sec.~\ref{sec:gamma_c}, sharply separates the dynamically stable and unstable regimes.
Below $\gamma_c$, the wave-packet dynamics quantitatively reproduces the stationary scattering prediction; at $\gamma_c$, the spectral singularity manifests as persistent lasing emission; above $\gamma_c$, an exponentially growing bound state dominates the dynamics and the stationary scattering coefficients lose physical relevance.
We stress that the smooth, finite $T(k)$ in panel~(c) gives no hint of the instability: the transition is invisible to the stationary coefficients themselves and is revealed only by the pole structure or by the time evolution.
An \textit{a priori} analysis of the $S$-matrix pole structure is therefore indispensable for assessing the dynamical stability of a non-Hermitian scattering system~\cite{zheng2025a}.

%%%%%%%%%%%%%%%%%%%%%%%%%%%%%%%%%%%%%%%%%%%%%%%%%%%%%%%%%%%%%%%%%%%%%%%%%%%%%%%%%%%%%%%%%%%%%%%%%%%%
\section{Implications for stationary scattering predictions}
\label{sec:implications}
%%%%%%%%%%%%%%%%%%%%%%%%%%%%%%%%%%%%%%%%%%%%%%%%%%%%%%%%%%%%%%%%%%%%%%%%%%%%%%%%%%%%%%%%%%%%%%%%%%%%

The stability boundary $\gamma_c(N) = 2\sin[\pi/(4N)]$ established and verified above sets the regime in which the stationary scattering picture is dynamically meaningful.
Since the stationary coefficients describe the actual dynamics only for $\gamma < \gamma_c(N)$, predictions made within the stationary framework must be read against this boundary.
Enlarging the periodic structure is the standard route to a richer scattering response: more unit cells introduce additional Fabry-P\'erot resonances, generate multiple CPA-laser points, and strengthen gain-loss-induced localization.
Yet the same enlargement lowers $\gamma_c$ and narrows the stable window.
We now revisit three phenomena studied extensively within the stationary framework, namely gain-loss-induced localization (Sec.~\ref{sec:localization}), reflectionless transport (Sec.~\ref{sec:anisotropic}), and CPA lasers (Sec.~\ref{sec:CPA_laser}), asking in each case which predicted features survive the stability constraint.

%%%%%%%%%%%%%%%%%%%%%%%%%%%%%%%%%%%%%%%%%%%%%%%%%%%%%%%%%%%%%%%%%%%%%%%%%%%%%%%%%%%%%%%%%%%%%%%%%%%%
\subsection{Gain-loss-induced localization}
\label{sec:localization}
%%%%%%%%%%%%%%%%%%%%%%%%%%%%%%%%%%%%%%%%%%%%%%%%%%%%%%%%%%%%%%%%%%%%%%%%%%%%%%%%%%%%%%%%%%%%%%%%%%%%

The Bloch-like index $\mu$ defined by the dispersion relation~\eqref{eq:mu_dispersion} classifies stationary transport through the periodic structure into three regimes~\cite{vazquez-candanedo2014,vazquez-candanedo2015,nguyen2016,achilleos2017a,moreno-rodriguez2020,lee2023,lazo2023,moreno-rodriguez2024}.
For energies satisfying $E^2 + \gamma^2 < 4$, the index $\mu$ is real, and the transmission coefficient $T$ [Eq.~\eqref{eq:T_exact}] oscillates as a function of $N$.
At the band edge $E^2 + \gamma^2 = 4$, one has $\mu = 0$ and $T = 1$ for all $N$.
For $E^2 + \gamma^2 > 4$, the index becomes purely imaginary, $\mu = i\phi$ with $\phi > 0$, and $T$ decays exponentially as $e^{-4N\phi}$ for large $N$~\cite{vazquez-candanedo2014}.
This exponential suppression of transmission, driven solely by balanced gain and loss without any disorder, has been termed gain-loss-induced localization~\cite{vazquez-candanedo2014}.

\begin{figure}
\centering
\includegraphics[width=\columnwidth]{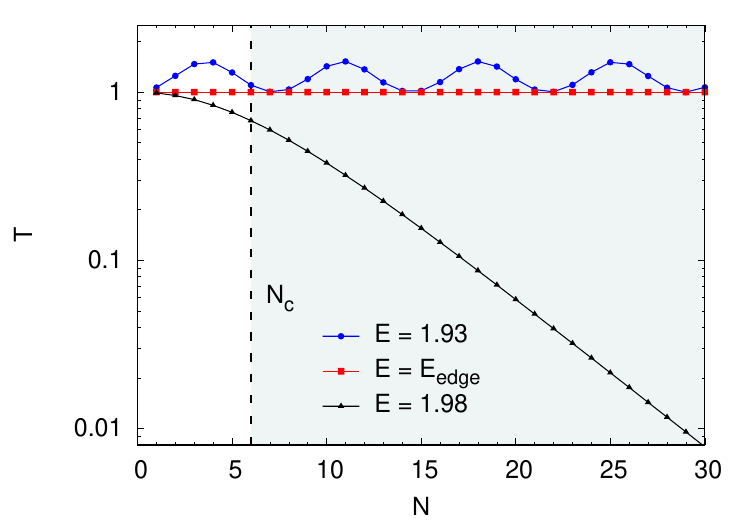}
\caption{
Transmission coefficient $T$ as a function of system size $N$ at $\gamma = 0.3$.
The parameter corresponds to the horizontal red dashed line in Fig.~\ref{fig:03_onset_condition}.
The blue circles ($E = 1.93$), red squares ($E = E_{\mathrm{edge}} = \sqrt{4-\gamma^2} \approx 1.977$), and black triangles ($E = 1.98$) represent the cases where the Bloch-like index $\mu$ is real, zero, and imaginary, respectively.
The vertical dashed line indicates the critical system size $N_c = 6$, beyond which the system becomes dynamically unstable (shaded region).
}
\label{fig:05_T_vs_N}
\end{figure}

Figure~\ref{fig:05_T_vs_N} illustrates these three regimes at $\gamma = 0.3$.
For $E = 1.93$, the Bloch-like index is real ($\mu \approx 0.217$), and $T$ oscillates as a function of $N$ with quasiperiod $\pi/(2\mu) \approx 7.2$ unit cells.
At the band edge $E = \sqrt{4 - \gamma^2} \approx 1.977$, the index vanishes and $T = 1$ for all $N$.
For $E = 1.98$, the energy lies in the evanescent regime ($\mu = i\phi$, $\phi \approx 0.051$) and $T$ decreases exponentially with $N$.

All three behaviors, however, are subject to the stability condition.
The parameters lie on the horizontal red dashed line at $\gamma = 0.3$ in Fig.~\ref{fig:03_onset_condition}, for which the threshold is crossed at the critical system size $N_c = 6$.
For $N \leq 5$, no first-quadrant pole exists, and the Bloch-like-index classification remains valid; for $N \geq 6$, at least one TGBS is present, and the system is dynamically unstable, irrespective of whether $\mu$ is real, zero, or imaginary [shaded region in Fig.~\ref{fig:05_T_vs_N}].
In the unstable regime, the exponentially growing bound state dominates the time evolution, so the stationary transmission, including the gain-loss-induced localization at $E = 1.98$, no longer reflects the wave-packet dynamics.

In Hermitian periodic systems, the Bloch-like index alone governs transport, and increasing $N$ simply extends the oscillatory or evanescent behavior indefinitely.
In periodic $\mathcal{PT}$-symmetric systems, increasing $N$ simultaneously lowers $\gamma_c \approx \pi/(2N)$, so the regime in which the Bloch-like-index classification retains physical relevance shrinks as the system grows.
This size-dependent stability constraint has no Hermitian counterpart and is fundamental to transport analysis in periodic non-Hermitian scattering systems~\cite{vazquez-candanedo2014,vazquez-candanedo2015,nguyen2016,shramkova2016,achilleos2017a,ge2017,wu2019a,wu2019b,zheng2019,moreno-rodriguez2020,lee2023,perez-garrido2023,guo2023a,lazo2023,lee2024,moreno-rodriguez2024}.

%%%%%%%%%%%%%%%%%%%%%%%%%%%%%%%%%%%%%%%%%%%%%%%%%%%%%%%%%%%%%%%%%%%%%%%%%%%%%%%%%%%%%%%%%%%%%%%%%%%%
\subsection{Reflectionless transport}
\label{sec:anisotropic}
%%%%%%%%%%%%%%%%%%%%%%%%%%%%%%%%%%%%%%%%%%%%%%%%%%%%%%%%%%%%%%%%%%%%%%%%%%%%%%%%%%%%%%%%%%%%%%%%%%%%
Periodic $\mathcal{PT}$-symmetric structures have been widely proposed as a platform for engineering reflectionless transport~\cite{vazquez-candanedo2014,vazquez-candanedo2015,ge2012,shramkova2016,achilleos2017a,wu2019a,zheng2019,moreno-rodriguez2020,lee2023,perez-garrido2023,lazo2023,lee2024}.
The generalized conservation relation~\eqref{eq:generalized_conservation} implies that whenever $T = 1$, the product $R_LR_R$ must vanish.
If only one reflection coefficient vanishes, the system is unidirectionally reflectionless; if both vanish simultaneously, it is bidirectionally reflectionless.

From the exact transmission coefficient [Eq.~\eqref{eq:T_exact}], perfect transmission $T=1$ arises through two distinct mechanisms.
The first is the band-edge condition $\mu = 0$ (i.e., $E^2 + \gamma^2 = 4$), already encountered in Sec.~\ref{sec:localization}.
At these energies, $T=1$ holds for any $N$.
The band-edge condition further implies $2\sin k = \gamma$, so the factor $(2\sin k - \gamma)$ in $M_{12}$ [Eq.~\eqref{eq:M12}] vanishes and $R_R = 0$.
Meanwhile, the factor $(2\sin k + \gamma) = 2\gamma$ in $M_{21}$ [Eq.~\eqref{eq:M21}] is nonzero, and the $\mu\to 0$ limit yields $R_L = 4N^2\gamma^2$.
The system is therefore unidirectionally reflectionless from the right.
This is the anisotropic transmission resonance identified in Ref.~\cite{ge2012}.

The second mechanism is the Fabry-P\'erot-like resonance condition~\cite{wu2019a,lee2023}
\begin{equation}
  \sin(2N\mu) = 0,
  \label{eq:BR_condition}
\end{equation}
which yields $N-1$ solutions within the propagating band,
\begin{equation}
  \mu_m = \frac{m\pi}{2N}, \quad m = 1,2,\ldots,N{-}1,
  \label{eq:BR_mu}
\end{equation}
at energies~\cite{vazquez-candanedo2014}
\begin{equation}
  E_{\text{FP}} = \pm\sqrt{4\cos^2\!\!\left(\frac{m\pi}{2N}\right) - \gamma^2}.
  \label{eq:BR_energy}
\end{equation}
Each solution requires $4\cos^2(m\pi/2N) > \gamma^2$ for $E_{\text{FP}}$ to be real, so the number of Fabry-P\'erot resonances decreases with increasing $\gamma$.
Since $M_{12}$ and $M_{21}$ [Eqs.~\eqref{eq:M12} and~\eqref{eq:M21}] are both proportional to $\sin(2N\mu)$, both reflection coefficients vanish at these energies, producing bidirectionally reflectionless transport~\cite{wu2019a,lee2023}.
Unlike the band-edge condition, the Fabry-P\'erot resonances exist only for $N \geq 2$, and their number grows linearly with $N$.

\begin{figure}
\centering
\includegraphics[width=\columnwidth]{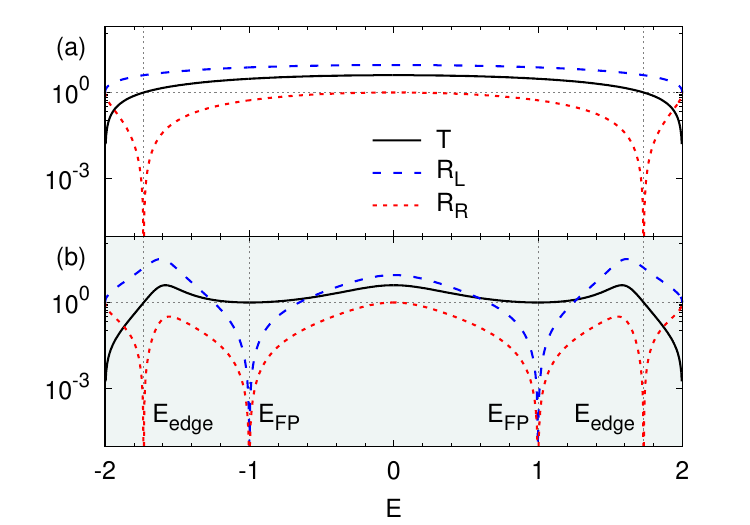}
\caption{
Scattering coefficients $T$ (solid black), $R_L$ (dashed blue), and $R_R$ (dotted red) as functions of $E$ at $\gamma = 1$ for (a) $N=1$ and (b) $N=2$.
The parameters correspond to the blue square and red triangle in Fig.~\ref{fig:03_onset_condition}, respectively.
Vertical dashed lines mark the band-edge energies $E = E_{\mathrm{edge}} = \pm\sqrt{3}$ in both panels, where $R_R = 0$ but $R_L \neq 0$.
In~(b), additional vertical dashed lines mark the Fabry-P\'erot-like resonance energies $E = E_{\text{FP}} = \pm 1$ [Eq.~\eqref{eq:BR_energy}], where both $R_L$ and $R_R$ vanish.
The shaded background in~(b) indicates that the system is dynamically unstable.
}
\label{fig:06_T_vs_E}
\end{figure}

Figure~\ref{fig:06_T_vs_E} illustrates both mechanisms at $\gamma = 1$.
For $N = 1$ [panel~(a)], only the band-edge condition contributes, giving $T=1$ with $R_R = 0$ at $E = E_{\mathrm{edge}}=\pm\sqrt{3}$.
For $N = 2$ [panel~(b)], an additional pair of Fabry-P\'erot resonances appears at $E = E_{\text{FP}}=\pm 1$, where both $R_L$ and $R_R$ vanish.

These two cases differ qualitatively when judged against the stability condition.
The parameters $(N, \gamma) = (1, 1)$ and $(2, 1)$ are marked by the blue square and red triangle in Fig.~\ref{fig:03_onset_condition}.
For $N = 1$, $\gamma_c = \sqrt{2} \approx 1.414$ and $\gamma = 1$ lies in the stable regime, so the predicted unidirectional reflectionlessness at $E = \pm\sqrt{3}$ is dynamically meaningful.
For $N = 2$, $\gamma_c = 2\sin(\pi/8) \approx 0.765$ and $\gamma = 1$ already exceeds this threshold [shaded region in Fig.~\ref{fig:06_T_vs_E}(b)].
A TGBS is present and the system is unstable, so neither the unidirectional reflectionlessness at $E = \pm\sqrt{3}$ nor the bidirectional reflectionlessness at $E = \pm 1$ describes the actual dynamics.
Proposals exploiting reflectionless transport must therefore confirm $\gamma < \gamma_c(N)$ before the predicted behavior can be physically realized.

%%%%%%%%%%%%%%%%%%%%%%%%%%%%%%%%%%%%%%%%%%%%%%%%%%%%%%%%%%%%%%%%%%%%%%%%%%%%%%%%%%%%%%%%%%%%%%%%%%%%
\subsection{CPA lasers}
\label{sec:CPA_laser}
%%%%%%%%%%%%%%%%%%%%%%%%%%%%%%%%%%%%%%%%%%%%%%%%%%%%%%%%%%%%%%%%%%%%%%%%%%%%%%%%%%%%%%%%%%%%%%%%%%%%

In Sec.~\ref{sec:S_matrix_poles}, spectral singularities were identified as $S$-matrix poles on the real $k$ axis, where $M_{22} = 0$ and the scattering coefficients diverge.
The corresponding Siegert state carries purely outgoing waves at a real energy and describes self-sustained lasing.
For the $\mathcal{PT}$-symmetric chain, the relation $M_{11} = M_{22}^*$ forces $M_{11}$ to vanish simultaneously, satisfying the condition for coherent perfect absorption~\cite{chong2010}.
Since lasing and coherent perfect absorption at the same frequency define a CPA laser~\cite{longhi2010,chong2011}, every spectral singularity of the periodic $\mathcal{PT}$-symmetric chain is a CPA-laser point.
Equation~\eqref{eq:gamma_n} therefore predicts $N$ CPA-laser points at $k = \pi/2$, their number growing linearly with the system size~\cite{vazquez-candanedo2015,shramkova2016,achilleos2017a,ge2017,wu2019b,zheng2019,lee2023}.
As $N$ increases, the lowest threshold descends to $\gamma_c = 2\sin[\pi/(4N)]$ [Eq.~\eqref{eq:gamma_c}], suggesting CPA-laser action at progressively weaker gain/loss strengths.

\begin{figure}
\centering
\includegraphics[width=\columnwidth]{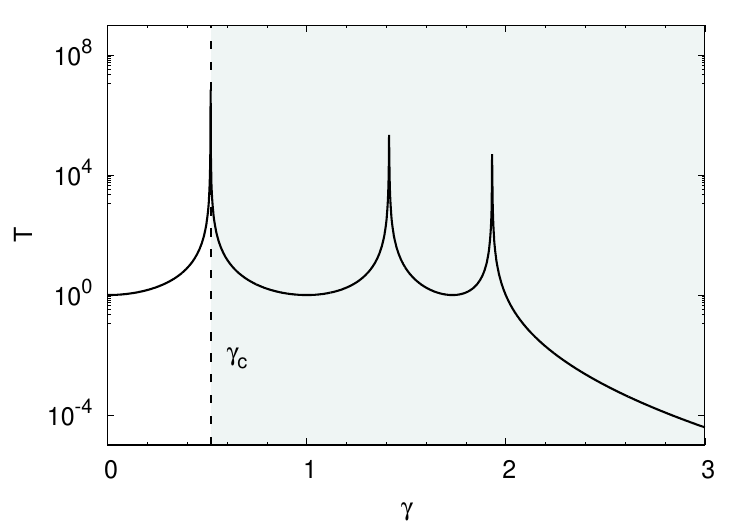}
\caption{
Transmission coefficient $T$ as a function of $\gamma$ for $N = 3$ at $k = \pi/2$.
The parameters correspond to the vertical dashed line at $N = 3$ in Fig.~\ref{fig:03_onset_condition}.
$T$ diverges at three CPA-laser points, corresponding to the spectral singularities shown in Fig.~\ref{fig:01_S_matrix_poles}(b), (e), and~(h), respectively.
The shaded region marks $\gamma > \gamma_c=(\sqrt{6}-\sqrt{2})/2$, where the system is dynamically unstable; only the first CPA-laser point, at $\gamma = \gamma_c$, is dynamically realizable.
}
\label{fig:07_T_vs_gamma}
\end{figure}

Of the $N$ CPA-laser points, however, only the one at $\gamma = \gamma_c$ is dynamically realizable.
The remaining $N-1$ points satisfy $\gamma_n > \gamma_c$ and lie inside the unstable regime, where the exponential growth of a TGBS overwhelms any lasing or absorption signature.
Figure~\ref{fig:07_T_vs_gamma} illustrates this for $N=3$ at $k=\pi/2$.
The transmission coefficient diverges at the three CPA-laser points [$\gamma = (\sqrt{6}-\sqrt{2})/2$, $\sqrt{2}$, and $(\sqrt{6}+\sqrt{2})/2$], corresponding to the spectral singularities in Fig.~\ref{fig:01_S_matrix_poles}(b), (e), and~(h).
Of these, only the lowest, $\gamma_c = (\sqrt{6}-\sqrt{2})/2$, falls on the stability boundary.
Consequently, each periodic $\mathcal{PT}$-symmetric chain admits a single dynamically realizable CPA-laser point, and the sole remaining design freedom is to lower its threshold by enlarging $N$.

%%%%%%%%%%%%%%%%%%%%%%%%%%%%%%%%%%%%%%%%%%%%%%%%%%%%%%%%%%%%%%%%%%%%%%%%%%
%%%%%%%%%%%%%%%%%%%%%%%%%%
\section{Summary and discussion}
\label{sec:conclusions}
%%%%%%%%%%%%%%%%%%%%%%%%%%%%%%%%%%%%%%%%%%%%%%%%%%%%%%%%%%%%%%%%%%%%%%%%%%
%%%%%%%%%%%%%%%%%%%%%%%%%%

In conclusion, we have analytically derived the instability threshold of a finite $\mathcal{PT}$-symmetric tight-binding chain of $N$ unit cells with balanced gain/loss strength $\gamma$. The instability is triggered by the emergence of TGBSs, which appear as $S$-matrix poles in the first quadrant of the complex $k$ plane. Analyzing this pole structure, we find that all spectral singularities of the periodic chain lie at the band center, which yields the closed-form threshold $\gamma_c = 2\sin[\pi/(4N)]$. For large $N$, this threshold scales as $\mathcal{O}(1/N)$ and vanishes in the thermodynamic limit. Time-domain wave-packet simulations quantitatively confirm the boundary: below $\gamma_c$, the dynamics faithfully follow the stationary scattering predictions; exactly at $\gamma_c$, the system reaches a lasing threshold and emits persistent radiation; and above $\gamma_c$, an exponentially growing TGBS emerges and completely overwhelms the stationary scattering picture.

Crucially, this scaling exposes a tension intrinsic to periodic $\mathcal{PT}$-symmetric systems: enlarging the structure is the standard route to sharpening the Bloch bands and enriching the stationary phenomenology, yet the same enlargement shrinks the stability window and drives the system unstable at progressively weaker gain and loss. Re-examined against this size-dependent boundary, several hallmarks of $\mathcal{PT}$-symmetric transport must be reassessed. For gain-loss-induced localization, larger $N$ deepens the exponential suppression of transmission but simultaneously lowers $\gamma_c$, so beyond a critical size the Bloch-like-index classification loses its physical relevance. For reflectionless transport, increasing $N$ multiplies the Fabry-P\'erot resonances, yet the shrinking threshold pushes many of the predicted reflectionless points into the unstable regime. Finally, for CPA lasers, of the $N$ predicted CPA-laser points only the first, at $\gamma_c$, is dynamically realizable, while the remaining $N-1$ are preempted by the onset of exponential growth.

Because the threshold marks the abrupt onset of exponential field amplification, it should be directly observable as self-starting oscillations in platforms ranging from coupled optical waveguides to photonic lattices and electrical circuits. Mapping the critical gain/loss strength against $N$ in such systems would furnish a direct experimental test of the $\mathcal{O}(1/N)$ scaling law derived here. Moreover, once the system enters the unstable regime, the exponentially amplified field will inevitably drive the gain medium into its nonlinear response. Characterizing this saturated, nonlinear regime, which lies entirely beyond the reach of linear stationary scattering theory, is a natural and necessary next step for future work.

Ultimately, our findings show that the scattering response of a periodic $\mathcal{PT}$-symmetric system is not governed by its Bloch phase alone, but by an unavoidable interplay between stationary band theory and finite-size stability constraints. Because the transfer-matrix method and related time-independent techniques remain the standard tools for designing non-Hermitian devices, the $S$-matrix pole analysis demonstrated here provides a general and rigorous framework for locating stability boundaries, delimiting the validity of stationary predictions, and defining the true physical operating limits of non-Hermitian scattering systems~\cite{zheng2025a}.

%%%%%%%%%%%%%%%%%%%%%%%%%%%%%%%%%%%%%%%%%%%%%%%%%%%%%%%%%%%%%%%%%%%%%%%%%%%%%%%%%%%%%%%%%%%%%%%%%%%%
\begin{acknowledgments}
This work was supported by the Natural Science Foundation of the Jiangsu Higher Education Institutions of China, Grant No. 22KJB140009.
\end{acknowledgments}
%%%%%%%%%%%%%%%%%%%%%%%%%%%%%%%%%%%%%%%%%%%%%%%%%%%%%%%%%%%%%%%%%%%%%%%%%%%%%%%%%%%%%%%%%%%%%%%%%%%%
%\bibliographystyle{plain}
\bibliography{MyLibrary.bib}

\end{document}